\documentclass[aps,prl,twocolumn,showpacs,superscriptaddress,a4paper]{revtex4-1}
\usepackage{graphicx}
\usepackage{amssymb,amsmath}
\usepackage{epsfig}
\usepackage{txfonts}

\usepackage{amsmath,amssymb,amsfonts}
\usepackage{mathrsfs}
\usepackage{empheq}

\usepackage{color}
\usepackage[unicode,breaklinks]{hyperref}

\usepackage[unicode]{hyperref}
\hypersetup{
    pdftitle={Anomalous},
    pdfauthor={Ding},
    pdfsubject={Strong Coupling},
    colorlinks=true,
    linkcolor=blue,
    citecolor=blue,
    filecolor=black,
    urlcolor=blue
}
\def\urlprefix{}
   \def\url#1{}

\newcommand{\be}{\begin{equation}}
\newcommand{\ee}{\end{equation}}

\DeclarePairedDelimiter\abs{\lvert}{\rvert}

\begin{document}
\title{Revealing Strong Plasmon-Exciton Coupling Between Nano-gap Resonators and Two-Dimensional Semiconductors at Ambient Conditions}	

\author{Jian Qin}
\affiliation{State Key Laboratory of Modern Optical Instrumentation, College of Optical Science and Engineering, Zhejiang University, Hangzhou 310027, China}
\author{Zhepeng Zhang}
\affiliation{Department of Materials Science and Engineering, College of Engineering, Center for Nanochemistry (CNC), College of Chemistry and Molecular Engineering, Academy for Advanced Interdisciplinary Studies, Peking University, Beijing 100871, People’s Republic of China}
\author{Yu-Hui Chen}
\affiliation{MacDiarmid Institute for Advanced Materials and Nanotechnology, Dodd-Walls Centre for Photonic and Quantum Technologies, Department of Physics, University of Otago, PO Box 56, Dunedin 9016, New Zealand} 
\author{Yanfeng Zhang}
\affiliation{Department of Materials Science and Engineering, College of Engineering, Center for Nanochemistry (CNC), College of Chemistry and Molecular Engineering, Academy for Advanced Interdisciplinary Studies, Peking University, Beijing 100871, People’s Republic of China}
\author{Richard J. Blaikie}
\affiliation{MacDiarmid Institute for Advanced Materials and Nanotechnology, Dodd-Walls Centre for Photonic and Quantum Technologies, Department of Physics, University of Otago, PO Box 56, Dunedin 9016, New Zealand} 
\author{Boyang Ding}
\email{boyang.ding@otago.ac.nz}
\affiliation{MacDiarmid Institute for Advanced Materials and Nanotechnology, Dodd-Walls Centre for Photonic and Quantum Technologies, Department of Physics, University of Otago, PO Box 56, Dunedin 9016, New Zealand} 
\author{Min Qiu}
\email{minqiu@zju.edu.cn}
\affiliation{School of Engineering, Westlake University, 18 Shilongshan Road, Hangzhou 310024, People’s Republic of China}
\affiliation{Institute of Advanced Technology, Westlake Institute for Advanced Study, 18 Shilongshan Road, Hangzhou 310024, People’s Republic of China}

\date{\today}

\begin{abstract}
	
Strong coupling of two-dimensional semiconductor excitons with plasmonic resonators enables control of light-matter interaction at the subwavelength scale. Here we develop strong coupling in plasmonic nano-gap resonators that allow modification of exciton number contributing to the coupling. Using this system, we not only demonstrate a large vacuum Rabi splitting up to 163 meV and splitting features in photoluminescence spectra, but also reveal that the exciton number can be reduced down to single-digit level ($N<10$), which is an order lower than that of traditional systems, close to single-exciton based strong coupling. In addition, we prove that the strong coupling process is not affected by the large exciton coherence size that was previously believed to be detrimental to the formation of plasmon-exciton interaction. Our work provides a deeper understanding of storng coupling in two-dimensional semiconductors, paving the way for room temperature quantum optics applications.

\end{abstract}

\maketitle
\textbf{Introduction|} Two-dimensional (2D) transitional metal dichalcogenides (TMDCs), such as molybdenum disulfide (MoS$_\text{2}$) and tungsten disulfide (WS$_\text{2}$) have attracted tremendous attention recently\cite{Splendiani2010,Mak2010}. These semiconductor nanosheets, when thinned down to monolayers (MLs), become direct bandgap, hosting excitons having ultralarge binding energy\cite{Ye2014a,Chernikov2014a,He2014} and very high oscillator strength\cite{Mak2010,Li2014}, which arise from the strong coulomb interaction and reduced dielectric screening in atomically thin structures. As a result, excitons in TMDC MLs can be tighly bound even at room temperature, producing strong light absorption and photoluminescence (PL). Integrating TMDC MLs with an optical resonator enables fast energy exchange between electromagnetic (EM) resonances and semiconductor excitons, i.e. the strong light-matter interaction or strong couling, allowing the formation of half-light half-matter quasiparticles, known as polaritons. The strong coupling process  not only is of interest for fundamental quantum optics, e.g. Bose-Einstein condensation\cite{Kasprzak2006a} with superfluid characteritics, but also exhibits a great potential for many compelling applications, e.g. quantum computing\cite{Amo2010} and thresholdless semiconductor lasing\cite{Ye2015a,Wu2015a}. 

A large coupling strength together with low damping loss are needed to access many of the key features of strong coupling. One solution to enhance the coupling strength $g$ is reducing the mode volume $V$ of EM excitations, since $g\propto\sqrt{N/V}$, where $N$ is the exciton number contributing to the coupling process. Traditional photonic resonators, such as Fabry-Perot (FP)\cite{Liu2015e,Dufferwiel2015d,Liu2015e} and photonic crystal cavities\cite{Zhang2018a}, though having low damping loss, are incapable of compressing mode volumes below the diffraction limit, which restrains the further enhancement of coupling strength. In this context, plasmonic resonators, noble metal nanoparticles allowing the excitation of surface plasmons, can highly confine incident photons into subwavelength volumes, providing ultra-compact and robust platforms for the realization of strong coupling at room temperature\cite{Wang2016s,Liu2016,Zhao2016d,Wurdack2017,Boulesbaa2016d}. Recent studies have successfully demosntrated the strong plasmon-exciton coupling in 2D TMDC semiconductors at the single nanoparticle level\cite{Cuadra2018,Wen2017}, exhibiting great advantages of plasmonic resonators in enhancing coupling strength as compared to traditional cavity systems. For example, utilizing nano-gap resonators, where TMDC nanosheets (mono- or multiple-layers) are tightly sandwiched in the gap between a nanoparticle and a metal film, Kleemann et al.\cite{Kleemann2018} and Han et al.\cite{Han2018} have significantly improved the vacuum Rabi splitting $\hbar\Omega_\text{R}$ to above 140 meV, which is almost three times of that in FP-like systems\cite{Liu2015e}.

\begin{figure} [t]
	\includegraphics[width=3.4in]{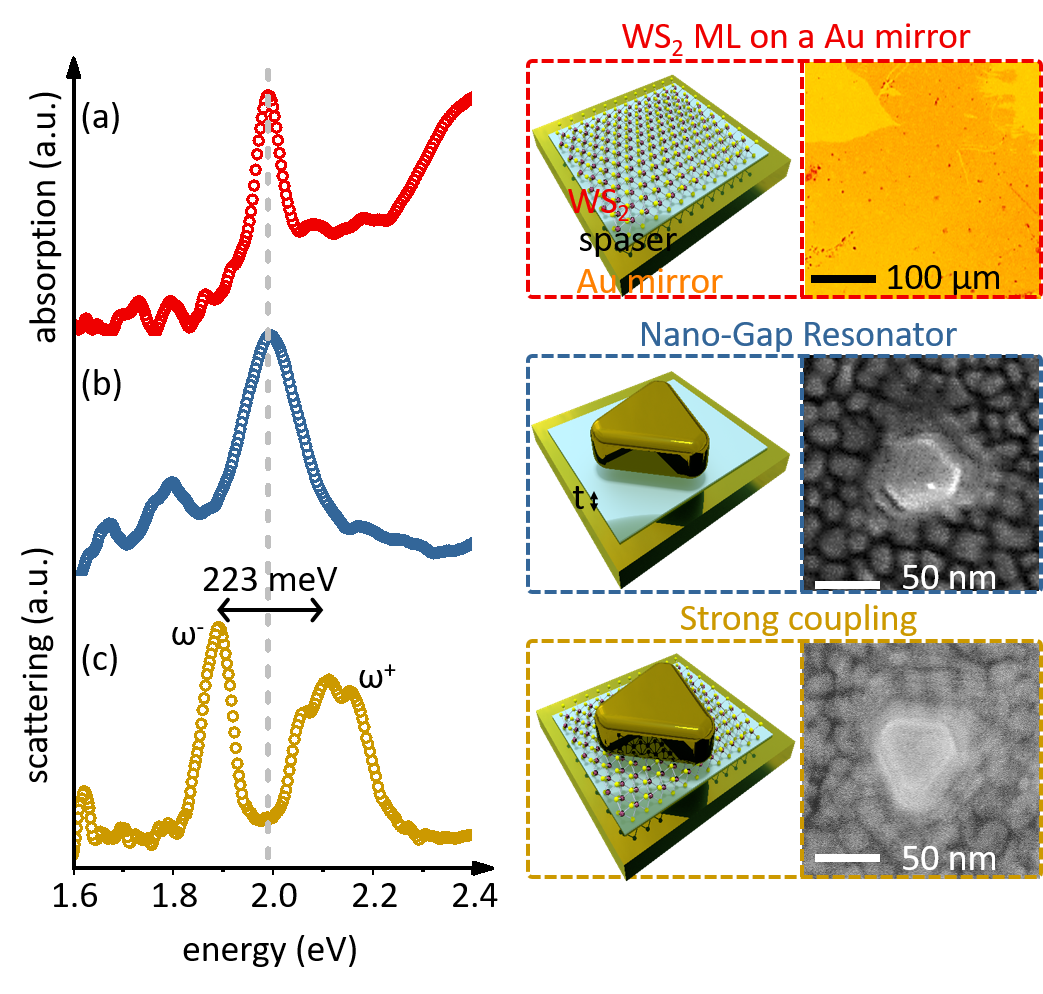}
	\caption{\textbf{Semiconductor excitons, plasmonic nano-gap resonators and strong coupling between them} (a) absorption spectrum of a WS$_{2}$ monolayer deposited on a gold film with a dielectric spacer that is as thick of $t \approx 1$\,nm; (b) DF scattering spectrum of a plasmonic nano-gap resonator that comprises a Au nanoprism separated from a Au mirror by a spacer with $t \approx 3$\,nm; (c) scattering spectrum of the WS$_\text{2}$ monolayer embeded in the nano-gap resonator with a spacer thickness $t \approx 1$\,nm. Middle (right) panels show the schematics (optical or scanning electron microscopes). }
	\label{F1}  
\end{figure}

To date, however, there remain a plenty of ambiguities in the development of plasmon-exciton coupling in TMDC MLs, mainly induced by unique excitonic properties in these crystalised ultrathin nanosheets. For example, it is unknown whether or not the exciton number $N$ can be flexibly tuned, because excitons in TMDC MLs are quasiparticles formed in semiconductor bandgap, unlike in the case of dye molecule\cite{Ni2008b,Balci2016} and rare-earth ion\cite{Zhao2013c,Lu2014b,Liu2017g} systems, where $N$ can be adjusted by changing the doping concentration; and it is uncertain whether or not the large exciton coherence size hinders further enhancement of coupling strength, since it is believed that EM dipoles must have larger dimensions than exciton coherence size to enable the coupling\cite{Kleemann2018}, which conflicts with the small mode volume of plasmonic resonators. In addition, it is still unclear how to effectively couple with excitons in TMDC MLs, which acquire complete in-plane dipole orientation\cite{Prins2014,Li2014,Wang2017a}. All these factors highly impede the realisation and control of coupling and prevent strong coupling from being realised at the fundamental limit of plasmonic resonators\cite{Ciraci2012,Chikkaraddy2016c}, thus significantly hampering the advances of promised applications.

Here we demonstrate room temperature strong plasmon-exciton coupling in nanoprism-film gap resonators that comprise gold (Au) nanoprisms and metal films sandwiching a WS$_\text{2}$ monolayer (Fig.~\ref{F1}). This system not only offers small mode volume $V$ that leads to a large average Rabi splitting up to 163 meV, but also, more importantly, allows modification of exciton number $N$ by adjusting the spacing between semicondutor MLs and the metal film. We find that the exciton number contributing to coupling in our systems can be reduced down to $N<10$, which is very close to the demand of quantum optics applications\cite{Hennessy2007a,Faraon2008}, which require involvement of a single exciton. In addition, we also note that the dimensions of EM dipoles in the structure do not exceed the exciton coherence diameter, indicating that the role of exciton coherence size is not decisive in enabling the strong coupling. Furthermore, we can uni-directionally excite the strong coupling features, enabled by the polarisation-insensitive eigenmodes in specially designed resonators. Finally, we show splitting features in PL spectra, possibly induced by PL coupling to lower polaritons and uncoupled excitons. Our work develops a convenient measure to engineer the mode hybridization, provideing a deeper understanding of strong coupling in 2D semiconductors and  potentially an ultra-compact platform for quantum exciton-polariton devices.

\textbf{Results~|}
Turning to the details of experiments, the WS$_2$ monolayer placed on a Au mirror with a dielectric spacer displays a narrow absorption peak at $ \hbar\omega_\text{0}=1.99 $\, eV, [Fig.\ref{F1}(a)] , corresponding to the exciton A (X$_\text{A}$ ) in  WS$_2$ MLs. The frequency of X$_\text{A}$ is slightly red-shifted as compared to previous literatures\cite{Shi2013,Chernikov2014a,Li2014}, induced by strain modifications during substrate transfer\cite{McCreary2016b,Krustok2017a}. [Fig.S4 in supplementary information (SI)] The schematic in Fig.~\ref{F1}(b) clearly shows the physical configuration of our nano-gap resonators, comprising a Au nanoprism separated from a Au film with a dielectric spacer. Upon light illumination, surface plasmons can be excited in the gap between nanoprisms and the film, as the result of coupling between the nanoprism and its image dipole in the mirror\cite{Mock2008,Moreau2012,Ciraci2012,Chen2015,Chen2017d}. In our structure, the  spacer is made of polyelectrolyte (PE) films using the layer-by-layer deposition. This enables us to precisely control the spacer thickness $\textbf{t}$ by changing the PE layer number  (Table S1 in SI), allowing the flexible tuning of gap resonance frequencies\cite{Mock2008,Chen2015}  (Fig.S1 in SI). For example, a $t\approx3\,\text{nm}$ spacer allows the excitation of gap resonances at a frequency identical to that of X$_\text{A}$, as shown in Fig.\ref{F1}(b). When a WS$_2$ ML is embedded in a nano-gap resonator that supports plasmon frequency matching the semiconductor exciton band [Fig.~\ref{F1}(c)] (Section 3 in SI for the discussion of frequency matching), the  hybrid system exhibits a typical splitting feature of strong coupling between plasmons and excitons, where the dark-field (DF) scattering spectrum shows two maixma at flanks of the excitonic energy of X$_\text{A}$, representing the lower ($\omega_-$) (LB) and upper ($\omega_+$) (UB) plasmon-exciton polariton branches. 

In a coupled hybrid system,  $\omega_{+}$ and $\omega_{-}$ are highly dependent on the detuning ($\delta=\omega_\text{pl}-\omega_\text{0}$) between exciton frequency $\omega_\text{0}$ and plasmon frequency $\omega_\text{pl}$. In our case, though $\omega_\text{pl}$ of each resonator is not tunable, we can analyse the plasmon-exciton coupling by studying a statistical splitting behaviours of a group of coupled nano-gap resonators that have different $\omega_\text{pl}$ [Fig.S6(a) in SI], using the Jaynes Cummings quantum mechanical model\cite{Bellessa2004,Agranovich2003}:
\begin{equation}
\label{dispersion01}
\omega_\pm=\dfrac{1}{2}(\omega_\text{pl}+\omega_\text{0})\pm\sqrt{g^2+\dfrac{\delta^2}{4}}
\end{equation}
where $g$ is the coupling strength. As the result, an average vacuum Rabi splitting $\hbar\cdot\Omega_{\text{R}} = \hbar\cdot 2g$ of ~163 meV is observed at the crossing point. This is far beyond the strong coupling criteria $g\geqslant 0.01\omega_\text{0}$ or $g>(\kappa-\gamma)/2$\cite{Schneider2018b} used in other works, where $\hbar\gamma = 50\,\text{meV}$  (excitonic linewidth) and $\hbar\kappa = 180\,\text{meV}$ (average linewidth of gap modes), which also outperforms that in other 2D semiconductor coupled plasmonic systems, including silver (Ag) nanoprisms\cite{Cuadra2018}, Ag nanorods\cite{Wen2017} and other types of plasmonic gap resonators\cite{Han2018,Kleemann2018}.

\textbf{Discussions~|}
Here we take a closer look at this significant mode splitting. As mentioned earlier, conventional treatments mainly focus on reducing the mode volume $V$ of resonators to achieve large coupling strength $g$, since  
\begin{equation}
\label{CS01}
g=\sqrt{N}\mu_\text{T}|E_\text{vac}|=\mu_\text{T}\,\abs*{\sqrt{\dfrac{N\cdot\hbar\omega}{2\epsilon\epsilon_\text{0}V}}}
\end{equation}
where $\mu_\text{T}$ is the transition dipole moment of excitons\cite{Yoshie2004}. Following Eq.~\eqref{CS01}, we should note that $g$ is also a direct reflection of the number $N$ of excitons involved in the coupling process. From application points of view, it is higly desirable to have a system that allows flexible tuning of $N$, as this will greatly enhance our control over the coupling process, yielding hybrid systems for various specific applications.  In our case, excitons are quasiparticles formed in semiconductor bandgaps and the number of atoms embedded within the gap is fixed, which makes it very hard to change $N$ by regular means.

\begin{figure} [b]
	\includegraphics[width=3.5in]{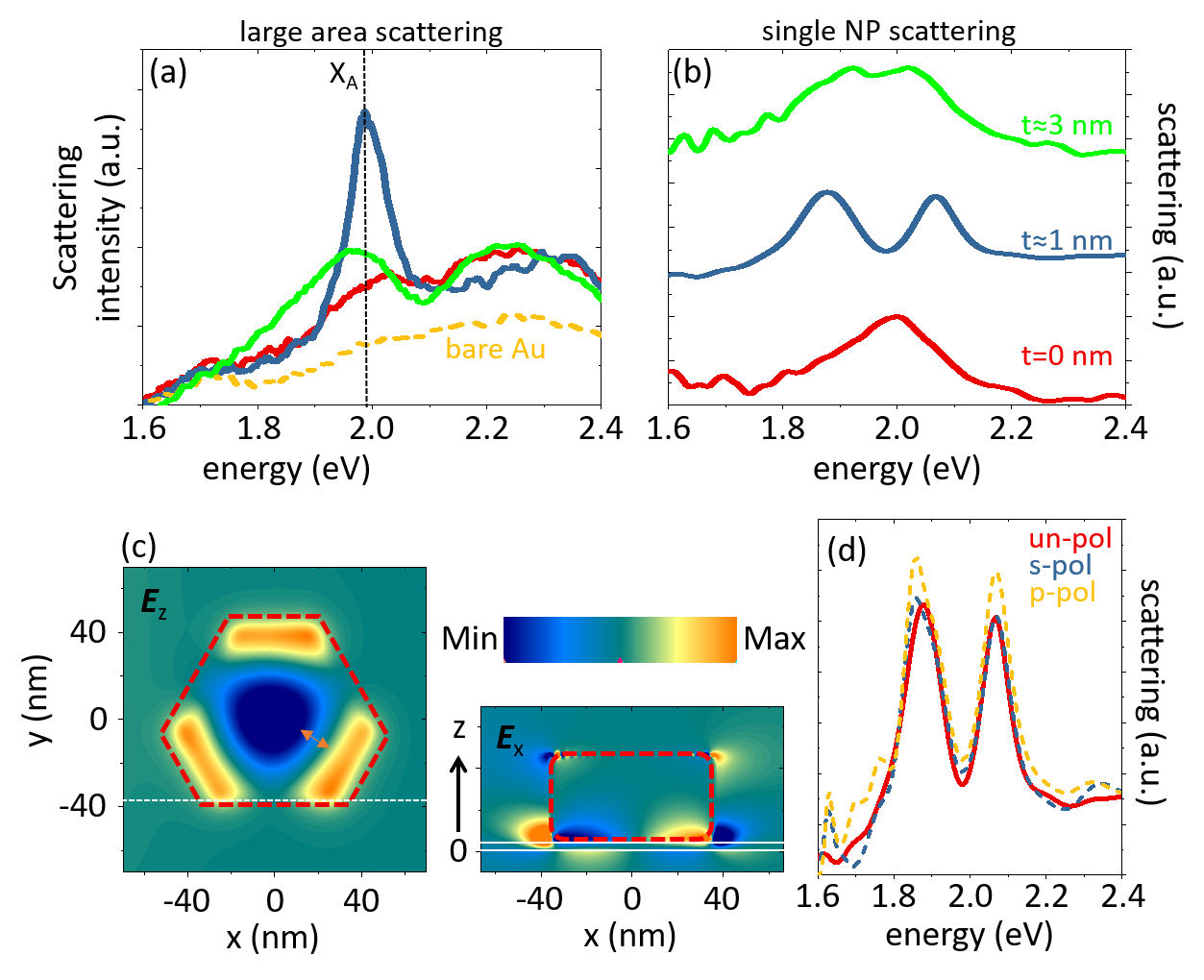}	
	\caption{\textbf{Modified exciton properties and unidirectional excitation} (a) scattered intensity from a large area ($>100\mu m^2$) of WS$_\text{2}$ MLs on a Au film with a $t=0\,\text{nm}$ (red), $t\approx1\,\text{nm}$ (blue) and $t\approx3\,\text{nm}$ (green) PE layer spacer under white-light illumination, with scattering from a bare Au film (yellow) being as a reference; (b) scattering spectra of nano-gap resonators embedded with WS$_\text{2}$ MLs with differently thick spacers having identical colour codes to (a); (c) z- (left) and x- component of electric fields viewed from x-y (within the gap) and x-z (crosss along white dashed line in the left panel)  planes at $E = 1.99\,\text{meV}$ corresponding to gap resonances; (d) scattering spectra of an identical nano-gap plasmonic resonator coupled with a WS$_\text{2}$ ML illuminated by white-light with different polarizations.}
	\label{F2}
\end{figure} 

\emph{Exciton Number\,|} Here we demonstrate that $N$ be can altered in our system and we use this property to analyse the coupling process. It is well known that the excitonic properties in TMDC MLs are extremely sensitive to surrounding EM environments\cite{Cheng2017a}, mainly because the atomically thin thickness makes the bandgap and excitons susceptible to Coulomb interaction\cite{Raja2017}. Fig.\ref{F2}(a) illustrates the white-light scattering collected from WS$_\text{2}$ MLs on Au substrates with different spacer thickness. In contrast with the reference from a bare Au film (yellow dashed), the scattering from $t\approx1\,\text{nm}$ sample (blue) gains an evident narrow peak at the X$_\text{A}$ frequency, while the scattering from the $t=0\,\text{nm}$ sample (red) only shows a barely identifiable maximum, indicating that the direct contact of WS$_\text{2}$ MLs on Au film significantly suppress the formation of the wannier-type excitons, induced by enhanced charge transfer from TMDCs to metals. In addition, if we slightly increase the spacer to $t\approx3\,\text{nm}$, the scattering (green) displays a broad and relatively red-shifted maximum (1.96 eV). (See Fig. S5 in SI for more details) These results agree very well with a previous work\cite{Cheng2017a}, which shows that the exciton strength in WS$_\text{2}$ MLs is negligible when in direct contact with a metal film, but reaches its maximum at $\sim1$\,nm from the metal film and then rapidly decays as the ML-metal distance increases. 

This distance dependent property will undoubtedly affect the plasmon-exciton coupling. Fig.\ref{F2}(b) demonstrates the scattering spectra of single coupled resonators with different gap spacing. When the spacer is absent ($t\approx0\,\text{nm}$), the system shows no mode splitting at all, and the $t\approx3\,\text{nm}$ system shows a broadened spectrum with a very shallow splitting as compared to the pronounced coupling feature of the  $t\approx1\,\text{nm}$ system. Using Eq.~\eqref{dispersion01}, we can obtain an average Rabi splitting $\hbar\cdot\Omega_\text{R}\approx80\,\text{meV}$ for $t\approx3\,\text{nm}$ system, much smaller than that of the $t\approx1\,\text{nm}$ system (163 meV). (see Fig.S6 and S7 in SI for more details) Using Eq.~\eqref{CS01} with $\mu_\text{T}=56\,\text{D}$\cite{Wen2017} and mode volume $V$ calculated for $t\approx1\,\text{nm}$ and 3 nm systems (Eq.S(2) in SI), we can estimate the exciton numbers $N$ in these two systems, with $N_{1\text{nm}}\approx14$ and $N_{3\text{nm}}\approx9$. These results clearly indicate that the exciton number highly relates to the distance between semiconductor MLs and metal films, significantly affecting coupling strength in hybrid systems. Much more importantly, the exciton numbers involved in the coupling process is an order lower of those in dye\cite{Wers2015} and other 2D semiconductor systems\cite{Han2018}, approaching the single exciton level, which provids a very promising platform for realisation of single exciton based strong coupling that is highly required for a variety of potential quantum optics applications\cite{Hennessy2007a,Faraon2008}.

\emph{Exciton Orientation and Coherence Size\,|}
As mentioned earlier, excitons in WS$_\text{2}$ MLs possess completely in-plane dipole orientation, which makes it difficult to couple with nano-gap resonators that typically acquire resonating orientation perpendicular to the film plane ($E_\text{z}$).\cite{Nordlander2004,Mock2008} In our nanoprism-film resonators, the electric fields of gap resonances [Fig.\ref{F2}(c)] show in-plane components that oscillate along the exciton orientation within the gap, thus capable of being coupled with 2D excitons. Furthermore, the unique tridiagonal geometry of nanoprisms makes gap resonances insensitive to excitation polarisations. As the result, the split scattering features can always be observed from the hybrid systems [Fig.\ref{F2}(d)], irrespective of unpolarized, p-polarized and s-polarized excitations, which greatly improve the operability and practicality of the coupled hybrid system in many exciton-related applications, e.g. valley-polarized excitations\cite{Sun2017a,Dufferwiel2017b,Chen2017b} and PL emission enhancement.

\begin{figure} [t]
	\includegraphics[width=3.5in]{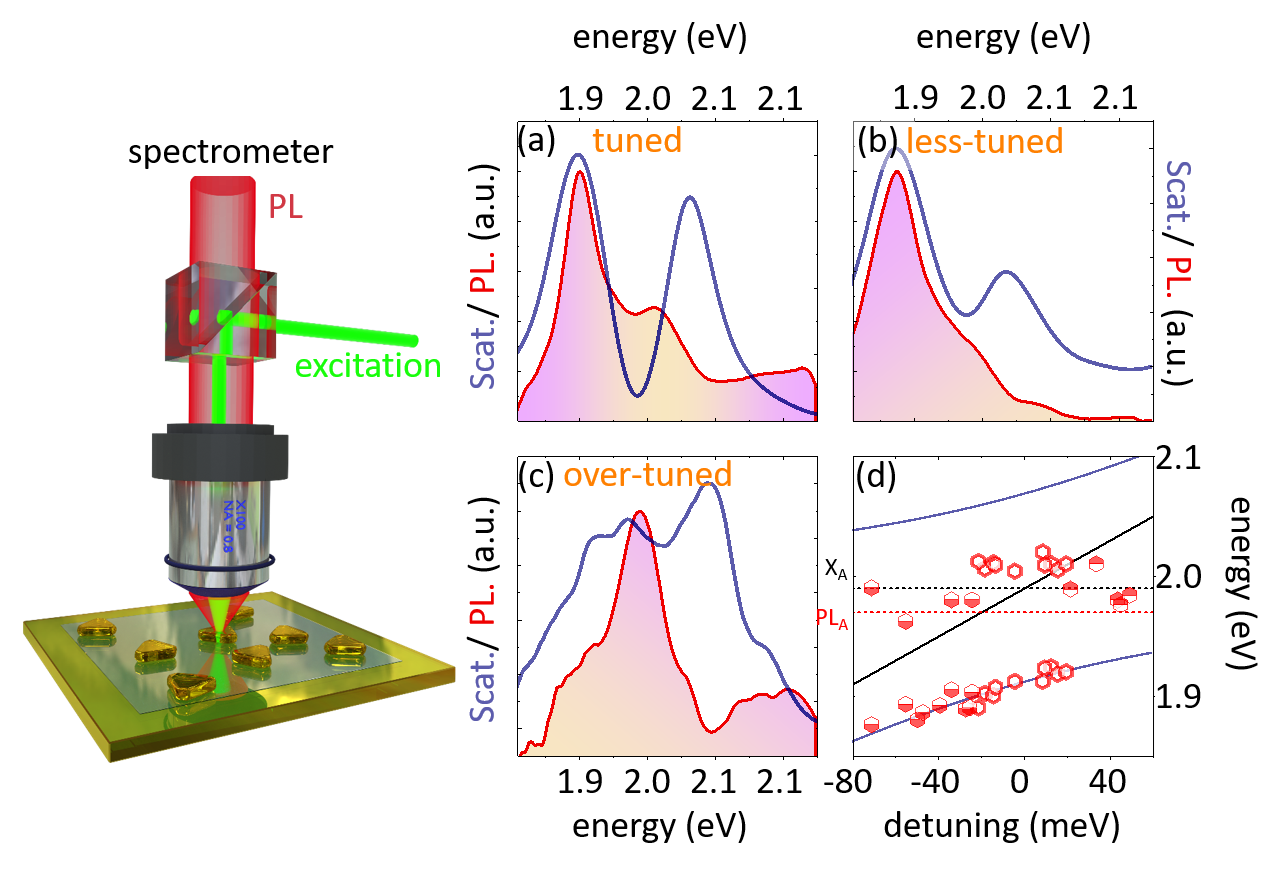}	
	\caption{\textbf{PL spectra in coupled hybrid systems with different tuning degrees.} PL (red) and scattering (cyan) spectrum for a group of nano-gap resonators ($t\approx1\,\text{nm}$) that stay tuned  (a),  less-tuned (b) and over-tuned (c); (d) spectral positions of PL maxima as a function of detuning $\delta$, where the upper-half solid, empty and lower-half solid hexgon symbols represent maxima frequency in over-tuned [$\hbar\omega_\text{pl}>(\hbar\omega_\text{0}+20\,\text{meV}$)],  tuned ($\hbar\omega_\text{pl}\approx\hbar\omega_\text{0}$) and less-tuned [$\hbar\omega_\text{pl}<(\hbar\omega_\text{0}-20\,\text{meV})$] PL spectra, respectively; blue and black curves indicate dispersion of plasmon-exciton polaritonic branches and gap resonances, respectively; Black (red) dashed horizontal lines indicate the spectral positions of X$_\text{A}$ absorption (PL).}
	\label{F3} 
\end{figure}

Here we comment on the role of exciton coherence size in the coupling process. The massive crystallised area ($\mu m$ size, see Fig.S5 in SI) enables high spatial extension of exciton delocalisation within TMDC MLs, resulting in large exciton coherence size, which, according to Ref.\cite{Kleemann2018}, requires EM dipoles with larger dimensions to couple with. This rule apparently conflicts with the needs of small mode volume $V$, restraining further enhancement of coupling strength $g$. However we note that this rule does not hold in our experiments. Specifically, following Ref.\cite{Feldmann1987}, we can obtain the exciton A coherence diameter in WS$_\text{2}$ MLs:
\begin{equation}
\label{CS02}
d_\text{c} = 4\,\sqrt{\dfrac{2h}{\gamma\cdot M}}
\end{equation}
using the homogeneous linewidth $\gamma$ of X$_\text{A}$ at room temperature\cite{Selig2016a} and the total exciton mass $M=m^*_\text{e}+m^*_\text{h}=0.59m_\text{e}$\cite{Shi2013}, we obtain the coherence diameter $d_\text{c}>30\,\text{nm}$. From Fig.\ref{F3}(c), it is clearly seen that the main component ($E_\text{z}$) of gap plasmons in $t\approx1\,\text{nm}$ system bear electric dipoles with opposite phase less than $20$\,nm apart from each other, which is smaller than $d_\text{c}$, but still can yield the splitting features. We therefore conclude that the exciton cohrerence size is not a decesive factor to enable the strong plasmon-exciton coupling.
\bigskip

\emph{Splitting in Photoluminescence\,|} Apart from DF scattering measurements, photoluminescence is another important observable to probe the coupled system, since exciton PL in TMDC monolayers not only relates to the bandgap structures, but also enables many significant applications, ranging from valley-polarized emission to solid-state lighting and displays. 

PL splitting in 2D semiconductors has been reported in studies using traditional resonators, such as FP-like cavities\cite{Liu2015e}. In plasmonic systems, however, only PL broadening or PL originating from the lower polariton\cite{Kleemann2018} were observed. This is possibly because plasmonic resonators, though having subwavelength dimensions, suffer from multimode resonances and high damping loss (large $\kappa$), which provides additional channels for non-radiative decay (phonons) to happen, thus considerably reducing the coupling cooperativity $C = g^2/(\gamma\cdot\kappa)$. As the result, the upper polariton branch is inactive due to the fast nonradiative energy transfer to lower electronic levels.

Here we demonstrate splitting features in PL spectra of our coupled nano-gap systems. Specifically, in the case of tuned system [Fig.\ref{F3}(a)], where $\omega_\text{pl}\approx\omega_\text{0}$, the PL spectrum shows two maxima, with one matching the scattering LB and another one peaking slightly above the excitonic frequency. For the less-tuned system ($\hbar\omega_\text{pl}<\hbar\omega_\text{0}$), the PL spectrum [Fig.\ref{F3}(b)] only shows one broadened maximum, coinciding with the LB of scattering, while the over-tuned system ($\hbar\omega_\text{pl}>\hbar\omega_\text{0}$) has only one PL peak [Fig.\ref{F3}(c)] close to the excitonic frequency with broadened spectral tails towards low energy. Fig.\ref{F3}(d) exhibits dispersions of PL maxima from many coupled systems as a function of detuning $\delta$. It is clear to see that there are two sets of maxima, with one dispersive set well matching LB polariton and another non-dispersive set slightly above the excitonic frequency. In our experiments, the non-dispersive branch can possibly be attributed to PL from uncoupled excitons\cite{Cuadra2017a}, given that most of the PL maxima's spectral positions are near the excitonic frequency.  The model of dark polariton states\cite{Herrera2017a} can also explain the non-dispersive PL branch. We have put relevant discussions into Fig.S8 in SI, but we should note that these uncommon PL behaviours originate from complicated interplay among plasmons, phonons and excitons and the specific mechanism is yet to be explored through more experimental and theoretical efforts.

\textbf{Conclusion\,|}  We have developed a hybrid system comprising a plasmonic nanoprism-film gap  resonator coupled with a WS$_\text{2}$ monolayer, which not only demonstrates an average vaccum Rabi splitting up to 163 meV, but also, more importantly, allowing modification of exciton number involved in the coupling process. Using this property to analyse the plasmon-exciton interaction, we show that the exciton number contributing to the coupling can be reduced down to single-digits ($N<10$). We also note that the exciton coherence size may not be relevant to the initiation of strong plasmon-exciton coupling. Furthermore, we find that the unique geometry of nanoprism-film resonator allows unidirectional excitation of the strong coupling observation in the scattering spectra. Finally, we have analysed the spectral splitting in photoluminescence, and the non-dispersive PL branch is possibly due to the excitation of uncoupled excitons or dark polariton states.

\section*{Acknowledgments} 
This work was supported in part by the National Key Research and Development Program of China (No. 2017YFA0205700) and the National Natural Science Foundation of China (Nos. 61425023, 61235007, and 61575177). In addition, the authors acknowledge the New Idea Research Funding 2018 (Dodd-Walls Centre for photonic and quantum technologies) and Smart Ideas Funds 2018 by MBIE New Zealand.

\bibliography{NPoM-WS2-SC}

\end{document}